\documentstyle[aps,prl,amssymb,twocolumn]{revtex}
\begin{document}
\wideabs{
\title{Detecting an association between Gamma Ray and Gravitational
  Wave Bursts} 
\author{Lee Samuel Finn\cite{bya} and Soumya D. Mohanty\cite{byb}}
\address{Center for Gravitational Physics and Geometry, The
  Pennsylvania State University, University Park PA 16802}
\author{Joseph D. Romano\cite{byc}}
\address{Department of Physical Sciences, The University of Texas,
  Browsnville TX 78521}
\date{\today}
\maketitle
\begin{abstract}
  If $\gamma$-ray bursts~(GRBs) are accompanied by gravitational wave
  bursts~(GWBs) the correlated output of two gravitational wave
  detectors evaluated in the moments just prior to a GRB will differ
  from that evaluated at times not associated with a GRB. We can test
  for this difference independently of any model of the GWB signal
  waveform.  If we invoke a model for the GRB source population and
  GWB radiation spectral density we can find a confidence interval or
  upper limit on the root-mean-square GWB signal amplitude in the
  detector waveband.  To illustrate we adopt a simple, physically
  motivated model and estimate that initial LIGO detector observations
  coincident with 1000~GRBs could lead us to exclude, with 95\%
  confidence, associated GWBs with ${h}_{\text{RMS}}\gtrsim
  1.7\times10^{-22}$.  This result does not require the detector noise
  be Gaussian or that any inter-detector correlated noise be measured
  or measurable; it does not require advanced or {\em a priori\/}
  knowledge of the source waveform; and the limits obtained on the
  wave-strength improve with the number of observed GRBs.
\end{abstract}
}
\narrowtext

Gamma Ray Bursts (GRBs), which are known to lie at cosmological
distances, likely arise from shocks in a relativistic fireball that is
triggered by rapid accretion on to a newly formed black hole
\cite{meszaros98a}.  The violent formation of a black hole is likely
to produce a substantial gravitational wave burst~(GWB); thus, we
expect GRBs to be preceded by GWBs that are candidate sources for the
new gravitational wave (GW) detectors now under construction
\cite{abramovici92a,bradaschia92a}.

Proposed GRB progenitors include coalescing or merging binary systems
({\em e.g.,} neutron star or black hole binaries, white dwarf - black
hole mergers, Helium core - black hole mergers) and hypernovae or
collapsars ({\em e.g.,} failed SNe Ib and single or binary Wolf-Rayet
star collapse) \cite{meszaros98a}.  In fact, statistical evidence
points to at least three different subclasses of GRBs
\cite{mukherjee98a}; so, the actual progenitors may include these as
well as other systems.  Matched filtering (MF) --- the method at the
focus of most of the gravitational wave detection literature ---
requires detailed knowledge of the actual GWB waveform; consequently,
it cannot be used to detect the distinct GWB associated with a GRB.
Additionally, since GRBs occur at cosmological distances ({\em i.e.,}
$z\gtrsim1$), the signal-to-noise ratio~(SNR) of any individual GWB
burst will almost certainly be insufficient for a high confidence
detection with these new detectors.  Detection techniques other than
MF that aim to detect distinct GWBs will perform even worse.

Here we suggest an alternative method for detecting a GWB/GRB {\em
  association.}  If GWBs are associated with GRBs, the correlated
output of two GW detectors will be different in the moments
immediately preceding a GRB ({\em on-source\/}) than at other times,
not associated with a GRB ({\em off-source\/}).  A statistically
significant difference between on- and off-source cross-correlations
would support a GWB/GRB {\em association\/} and, by implication,
represent a detection of gravitational waves by the detector pair.
(While we focus on GRBs in this paper any plausible class of
astronomical events can serve as a trigger.)

We can test for a difference between the on- and off-source
cross-correlation populations using Student's $t$-test, {\em
without specifying an {\em a priori\/} model for the signal waveform,
source, or source population.} The observed $t$ statistic can also be
used to set a confidence interval (CI) or upper limit (UL) on the
root-mean-square (RMS) amplitude of GWB signals associated with GRBs,
where the average is over the source population.  If we specify even a
rudimentary GWB source model, the CI/UL constrains the model.

We restrict attention here to the two full-length LIGO detectors
(denoted ${\cal D}_i$, $i=1,2$).  These detectors are nearly
identically oriented and lie $\sim3000$~Km apart.  We do not require
that the detector noise statistics be Gaussian, although we do require
here that they be approximately stationary.  Without loss of
generality we assume the noise has zero mean and denote its {\em
one-sided\/} power spectral density (PSD) by $S_i(f)$.  Noise
cross-correlated between ${\cal D}_1$ and ${\cal D}_2$ is assumed only
to be stationary (but not necessarily Gaussian) and weak compared to
the intrinsic noise of both the detectors.

\paragraph{The on-source and off-source distributions.}
Suppose that a GWB, associated with a GRB, is incident from direction
$\vec{n}$ on the GW detector ${\cal D}_i$ at time $t_{a}^{(i)}$.  The
{\em lag\/} $\delta t$, equal to $t_{a}^{(2)}-t_{a}^{(1)}$, depends
only on $\vec{n}$, which we know from the GRB observation.  The lag is
also the same as the difference $t_\gamma^{(2)}-t_\gamma^{(1)}$, where
$t_\gamma^{(i)}$ is the arrival time at detector ${\cal D}_{i}$ of the
GRB.

Assuming that any associated GWB precedes the GRB, focus attention on
the output $x_i(t)$, for $0\leq t_\gamma^{(i)}-t\leq T$, of detector
${\cal D}_i$.  Choose $T$ as long, but no longer, than necessary to
insure that $x_i$ includes the possible GWB signal.  From the $x_i(t)$
we compute the weighted cross-correlation
\begin{eqnarray}
X&:=&\langle x_1,x_2\rangle\nonumber\\
&:=&\int\!\!\!\int_0^T \!\!\!\!dt\, dt^\prime\,
x_1(t_\gamma^{(1)}-t)
Q(|t-t^\prime|) x_2(t_\gamma^{(2)}-t^\prime).
\label{innpro}
\end{eqnarray}
The filter kernel $Q$ is at our disposal: we discuss its choice
in section~\ref{sec:Q} below.

The collection of $X$ computed for each of $N_{\text{on}}$ GRBs
form a set ${\cal X}_{\text{on}}$ of {\em on-source\/} events.  To
complement that set, we also construct a set ${\cal X}_{\text{off}}$
of $N_{\text{off}}$ {\em off-source events,} using data segments
$x_{i}$ corresponding to random sky directions and arrival times not
associated with any GRB.

The sample sets ${\cal X}_{\text{off}}$ and ${\cal X}_{\text{on}}$ are
drawn from populations whose distributions we denote $p_{\text{off}}$
and $p_{\text{on}}$.  For $T$ much greater than the detector noise
auto- and cross-correlation times, the central limit theorem
\cite{stuart94a} implies that $p_{\text{off}}$ is normal,
\begin{mathletters}
  \label{eq:poff}
  \begin{equation}
    p_{\text{off}}(X) \simeq N(X;\,\mu_{\text{off}},\sigma),
    \label{poffx}
  \end{equation}
  with mean and variance
  \begin{eqnarray}
    \mu_{\text{off}}&:=&\text{E}[\langle n_1,n_2\rangle],\\
    \sigma^2&:=&\text{E}\left[ \left( \langle n_1,n_2\rangle
        - \mu_{\text{off}}\right)^2\right].
  \end{eqnarray}
\end{mathletters}
Here $n_{i}(t)$ denotes noise from detector $i$ and $\text{E}[\cdot]$
represents an ensemble average across the detector output.  Note that
$\mu_{\text{off}}$ is just the detector noise cross-correlation
evaluated at the lag $\delta t$.  (A weak assumption behind equation
\ref{poffx} is that the noise cross correlation does not vary
significantly over the lag $\delta t$, which should be the case for
terrestrial noise sources.)

Now suppose that GRBs are preceded by GWBs. Elements of
${\cal X}_{\text{on}}$ then take the form
\begin{equation}
    X = 
    \left<n_{1},n_{2}\right>
    + \left<h_{1},n_{2}\right>
    + \left<n_{1},h_{2}\right>
    + \left<h_{1},h_{2}\right>,
\end{equation}
where $h_{i}(t)$ is detector $i$'s response to the incident GWB.
Define $P_{i}$ by
\begin{equation}
    P_{i} :=
    4\int_{0}^{\infty} df\,|\widetilde{h}_{i}(f)|^{2}/S_{i}(f).
\end{equation}
If $\overline{P_i}$, the average of ${P}_{i}$ over the source
population, is much less than unity, then $p_{\text{on}}$ is also a
normal distribution with variance $\sigma^2$ and mean
\begin{mathletters}
  \label{eq:pon}
  \begin{eqnarray}
    \mu_{\text{on}} &=& \mu_{\text{off}}+\overline{s},\qquad\text{where}\\
    s &:=& \left<h_{1},h_{2}\right>\label{eq:sdef}
  \end{eqnarray}
\end{mathletters}
and $\overline{s}$ is, again, an average of $s$ over the source
population.

\paragraph{Student's $t$-test.}\label{Sec:t}
Pose the null hypothesis
\begin{equation}
H_0:\; p_{\text{off}}(X) = p_{\text{on}}(X)\;.
\label{nullh}
\end{equation} 
Rejecting $H_0$ supports a GWB/GRB association.  Since $p_{\text{on}}$
and $p_{\text{off}}$ are normal and differ, if at all, only in their
means, we can test $H_0$ using Student's $t$-test \cite{snedecor67a}.

The $t$ statistic is defined from ${\cal X}_{\text{on}}$ and ${\cal 
X}_{\text{off}}$ by
\begin{mathletters}
\begin{eqnarray}
  t &:= &
  {\hat{\mu}_{\text{on}}-\hat{\mu}_{\text{off}}\over\Sigma}
  \sqrt{N_{\text{on}}N_{\text{off}}\over
    N_{\text{on}}+N_{\text{off}}},
  \label{tstatistic}  \\
  \Sigma^2 &=& {
    \hat{\sigma}^{2}_{\text{on}}\left(N_{\text{on}}-1\right) +
    \hat{\sigma}^{2}_{\text{off}}\left(N_{\text{off}}-1\right)
    \over
    N_{\text{on}}+N_{\text{off}}-2} ,
\end{eqnarray}
\end{mathletters}
where $\hat{\mu}_{\text{on}}$ and $\hat{\mu}_{\text{off}}$
($\hat{\sigma}_{\text{on}}^2$ and $\hat{\sigma}_{\text{off}}^2$) are
the {\em sample\/} means (variances) of ${\cal
  X}_{\text{on}}$ and ${\cal X}_{\text{off}}$, respectively.  

The expectation value of $t$, averaged over the source population and
across the detector noise processes, is
\begin{equation}
  \mu_t := E[t] = {\overline{s}\over\sigma}
  \sqrt{N_{\text{on}}N_{\text{off}}\over
    N_{\text{on}}+N_{\text{off}}}. \label{eq:def:mut}
\end{equation}
The relative orientation of the two LIGO detectors guarantees that
$h_1(t)$ and $h_2(t)$ are very nearly identical.  It follows that, for
LIGO, $\overline{s}$ is non-negative; correspondingly, the expectation
value $\text{E}[t]$ of $t$ is positive in presence of a GWB/GRB
association and zero otherwise.

The actual value of $t$, given any observed sets ${\cal
X}_{\text{on}}$ and ${\cal X}_{\text{off}}$, will vary from $\mu_t$. 
The distribution of $t$ is normal for large $N_{\text{on}} +
N_{\text{off}}$ and is, more generally, tabulated in any standard
statistics text \cite{stuart94a}.  Consequently, we can find a $t_0$
such that, when $H_0$ is true ($\mu_t=0$) $t$ is greater than $t_0$ in
less than a fraction $\alpha$ ({\em e.g.,} 5\%) of all observations. 
This is our test: if we observe $t$ greater than $t_{0}$ we reject
$H_{0}$ and conclude that we have found evidence of a GWB/GRB
association with significance $1-\alpha$ ({\em e.g.,} 95\%).


\paragraph{The filter kernel $Q$.}\label{sec:Q} 
If we knew the signal $h_{i}(t)$ corresponding to each GRB trigger we
could construct a $Q$ that maximizes $s$:
\begin{equation}
  {Q}(\tau) = 
  \int_{-\infty}^{\infty} df\,e^{2\pi if\tau}
  {\widetilde{h}_{1}(f)\widetilde{h}^{*}_{2}(f) \over 
    S_1(|f|)S_2(|f|)},
  \label{eq:Q}
\end{equation}
where $\widetilde{h}_i$ is the Fourier transform of $h_i$.  For the
LIGO detectors, the $h_{i}$ are identical. Denoting their common
functional form $h(t)$, the optimal $Q$ depends only on the GWB signal
spectral density $|\widetilde{h}(f)|^{2}$.

More generally, we can put any knowledge we have of the signal's
spectral density shape into $Q$. For LIGO we can choose $Q$ to match
the signal model irrespective of the details of the possible GWB
waveforms if $|\widetilde{h}(f)|^2$ is independent of signal
parameters. This happens, for instance, in the case of the quadrupole
formula waveform of an inspiraling binary.  For GWBs associated with
GRBs there is no reason to believe that $|\widetilde{h}(f)|^{2}$ will
be known {\em a priori,} let alone that it have this special property.
Lacking detailed knowledge, we recommend adopting $Q$ given by
equation \ref{eq:Q} with $|\widetilde{h}(f)|^2$ assumed to be unity in
the detector bandwidth.


\paragraph{Setting upper limits.}
Having specified $Q$ we can test $H_0$ as described above to rule on
the presence of a GWB/GRB association, {\em independently of any model
  for the GWB signal or its source.} Alternatively, we can use the
observed $t$ to establish a confidence interval (CI) or upper limit
(UL) on $\mu_t$, and hence $\overline{s}$, which is related to the GWB
wave-strength (cf.\ eq.~\ref{eq:def:mut}, \ref{eq:sdef}).  If we
invoke a model for the spectral density $|\widetilde{h}(f)|^2$
and spatial distribution of GWB/GRB sources, this becomes a
physical constraint on the model.

To obtain a CI/UL on $\overline{s}$ given an observed $t$, we
construct a {\em confidence belt} \cite{stuart94a}, of desired
confidence, in the $t$--$\mu_t$ plane.  We follow the construction of
\cite{feldman98a}, which unifies the treatment of CIs and ULs.
(\cite[table X]{feldman98a} tabulate UL/CIs 
appropriate to our case: the mean $\mu$ of a Gaussian variable $x$
when $\mu \geq 0$ and $x$ itself is the observation.)  From observed
sets ${\cal X}_{\text{on}}$ and ${\cal X}_{\text{off}}$ the
corresponding $t$ is computed.  From $t$ and the confidence belts the
corresponding CI/UL on $\overline{s}$ can be read-off.

To interpret the CI/UL, imagine an ensemble of observations and
corresponding pairs of sample sets ${\cal X}_{\text{on}}$ and ${\cal
X}_{\text{off}}$.  Corresponding to this ensemble of observations is
an ensemble of values $t$ and CI/ULs.  If the confidence level is
$\epsilon$ ({\em e.g.,} 95\%), then a fraction $\epsilon$ of these
CI/ULs will include the actual value of $\overline{s}$ (and a fraction
$1-\epsilon$ will not).  It is in this sense that the CI/UL
corresponding to the observed $t$ is said to bound $\overline{s}$ with
confidence $\epsilon$.

To measure the effectiveness of the proposed test we evaluate the UL
{\em most likely\/} to be placed on $\overline{s}$ if $H_0$ is, in
fact, true.  When $H_{0}$ is true the most likely {\em observed\/} $t$
is zero.  Denoting the corresponding UL on $\mu_t$ as $\mu_{t,\max}$
the UL on $\overline{s}$ is
\begin{mathletters}
\label{mutlim}
\begin{eqnarray}
{\overline{s}\over\sigma}&\leq&\mu_{t,\max}
\sqrt{N_{\text{on}}+N_{\text{off}}
\over N_{\text{on}}N_{\text{off}}}\\
&=& \left\{\begin{array}{ll}
\mu_{t,\max}\sqrt{2/N_\gamma}&(N_{\text{on}}=N_{\text{off}}=N_{\gamma})\\
\mu_{t,\max}/\sqrt{N_{\text{on}}}&(N_{\text{off}}\gg N_{\text{on}})
\end{array}\right.
\end{eqnarray}
\end{mathletters}
Since the duty cycle of GRBs is low, the size of the off-source sample
can be made much larger than the size of the on-source sample. Even if
both sample sets are the same size, however, the limit obtained will
be weaker by only a factor of $2^{1/2}$. 

The upper limit $\mu_{t,\max}$ corresponding to an observed $t$ of
zero and different degrees of confidence is given in \cite[table
X]{feldman98a}.  For reference we note that $\mu_{t,\max}$ is 1.00 for
68.27\%, 1.64 for 90\%, 1.96 for 95\%, and 2.58 for 99\% confidence.

A derived CI/UL on $\overline{s}$ implies, within the context of a
GWB/GRB source model, a CI/UL on the RMS GWB signal amplitude in the
detector band, with the average over the source population.  As an
example, suppose that each GRB is accompanied by the formation of a
several solar mass black hole and a corresponding millisecond
timescale GWB in the source rest frame.  Assume further that
$|\tilde{h}(f)|^{2}$ is approximately constant in the corresponding
KHz bandwidth $B_{s}$.  (This is consistent with numerical models of
supernova core collapse \cite{finn91a,monchmeyer91a} and with the
formation or ring-down of all but the most rapidly rotating solar mass
black holes \cite{finn92a}.)  At the detector, the signal power from a
source at redshift $z$ lies in the bandwidth $B_s/z'$, where $z'$
is equal to $1+z$.

For simplicity, assume that the detector noise PSDs $S_i(f)$ are
identical and equal to a constant $S_0$ in the detector bandwidth
$B_d$, which we take to be approximately 100~Hz about a central
frequency of 150~Hz.  Outside the detector band we set $S_i$ equal to
infinity.  (This is a rough approximation to the actual shape of the
noise PSD of LIGO \cite{LIGOE95001802}.)  Finally, note that $B_{s}$
is much larger than $B_{d}$, so that $B_s/z'$ completely overlaps
$B_d$ for some large range of $z'$.

With these assumptions, 
\begin{eqnarray}
s &=&  \int_{-\infty}^\infty df \, 
|\widetilde{h}(f)|^2 \widetilde{Q}(f) =
{2A^2 B_d\over S_0^2}
\qquad\text{and}\label{seg}\\
\sigma^2& = &{T\over4} \int_{-\infty}^\infty df\,
S_1(|f|)S_2(|f|) |\widetilde{Q}(f)|^2 = {T B_d\over2 S_0^2},
\label{seg2}
\end{eqnarray}
where $A$ is defined by
\begin{equation}
\int_{-\infty}^{\infty} df\ 
|\widetilde{h}(f)|^2 = {2 A^2 B_s\over z'}.
\label{aeff_def}
\end{equation}
From equations \ref{seg}, \ref{seg2} and \ref{aeff_def} it follows
that 
\begin{equation}
{\overline{s}\over\sigma} = 
\text{E}\left[ 
  {2\sqrt{2} A^2 B_d\over
  \sqrt{T B_d}S_0}
\right] \simeq  {2\sqrt{2}\ \overline{A^2}\
  B_d\over \sqrt{T B_d} S_0},
\label{avrho}
\end{equation}
where we have replaced $A^{2}$ by its mean over the source population
(a good approximation when $A$ is sharply peaked about its mean.)
From equations \ref{mutlim} and \ref{avrho} and assuming that $H_0$ is
true we find
\begin{equation}
\overline{A^{2}}\leq A_{\max}^2 = {\mu_{t,\max}\over2\sqrt{2}}
\left[T B_d\over N_{\text{on}}\right]^{1/2}
{S_0\over{B_d}},
\label{upplima0}
\end{equation}
with $N_{\text{off}}\gg N_{\text{on}}$ and $\mu_{t,\max}$ obtained
from the confidence belt construction \cite{feldman98a} with $t=0$.

We expect that different GWBs will have different waveforms
and durations. Define the RMS signal power in the detector band by 
\begin{equation}
  h^2_{\text{RMS}} := \overline{\left[ 
    {2\over \tau}\int_{f\in{}B_{d}}
    df \, |\tilde{h}(f)|^2
\right]}, \label{eq:hrms}
\end{equation}
where $h(t)$ is the GWB waveform, $\tau$ its duration in the detector
band, and the average is over the source population.  In our example
--- broadband bursts whose bandwidth includes the detector
band --- we can approximate $1/\tau$ by the detector bandwidth $B_d$. 
Combining equations \ref{eq:hrms}, \ref{upplima0} and \ref{aeff_def}
we find the UL on ${h}_{\text{RMS}}$:
\begin{eqnarray}
  {h}_{\text{RMS}}^2 &\leq& \left[1.7\times10^{-22}\right]^2
  {\mu_{t,\max}\over 1.96}
  \left(
    {T\over0.5\,\text{s}}{1000\over N_{\text{on}}}
  \right)^{1/2}\nonumber\\
  &&\qquad\times
  {S_0\over\left( 3\times10^{-23}\,\text{Hz}^{-1/2}\right)^2}
  \left({B_d\over 100\,\text{Hz}}\right)^{3/2}.
  \label{hrms1}
\end{eqnarray}
The reference values of $B_d$ and $S_0$ are characteristic of the
initial LIGO detectors \cite{LIGOE95001802}.  For $T$ (cf.\ 
eq.~\ref{innpro}) we assume GRBs are generated by internal shocks in
the fireball; then, the GRB/GWB delay is approximately $0.1$~sec in the
source rest frame \cite{rees94a}. To accommodate GRBs at redshifts
$z\leq4$ we take $T\sim0.5$~sec. Finally, $\mu_{t,\max}$ equal to
1.96 corresponds to a 95\% confidence UL \cite{feldman98a}.

If, on the other hand, GRBs are generated when the fireball is
incident on an external medium, then \cite[eq.\ 3.6]{meszaros93a} with
$n_1 = 1$, $\alpha=1$, $E_{51} = 10$, and $\Gamma\gtrsim100$ gives a
source rest-frame delay $\lesssim100$~sec, in which case $T$ should be
500~s and the corresponding UL on $h_{\text{RMS}}$ is
$9.4\times10^{-22}$. 

Two final notes are in order.  To calculate the $X$ (cf.\ 
eq.~\ref{innpro}), which are at the heart of our analysis, we must
know accurately the GRB source direction.  Bright bursts in the
BATSE3B catalog have positional accuracies of $\delta\theta\lesssim
1.5^\circ$ \cite{meegan96a}.  The corresponding uncertainty in $s$ is
$\lesssim5$\%, which does not affect significantly the UL on
$\overline{s}$.

Finally, the proposed BATSE follow-on --- SWIFT --- is not an all-sky
GRB detector.  It will have greater sensitivity than BATSE, but
observe only a fraction of the sky at any one time. If SWIFT pointing
favors the sky normal to the LIGO detector plane, LIGO's sensitivity
to GWBs from observed GRBs will be maximized, increasing the
sensitivity of the test described here.

\paragraph{Conclusions.}
Gamma-ray bursts (GRBs), which are believed to be associated with the
violent formation of a stellar mass black hole, may well be
immediately preceded by a gravitational wave burst (GWB).  If we
compare the correlated output of two gravitational wave detectors
immediately preceding a GRB to the correlation at other times, not
associated with a GRB, then a statistically significant difference is
evidence for a GRB/GWB association.

We can test for this difference --- independent of any model of the
GRB/GWB source or GWB waveform --- using Student's $t$-test.
Alternatively, we can set an upper limit (UL) or confidence interval
(CI) on the RMS GWB amplitude in the detector waveband, averaged over
the source population.  This CI/UL constrains any GRB/GWB model we do
invoke.

This analysis has several important advantages over matched filtering,
which is the method at the focus of most of the gravitational wave
detection literature.  In particular, it becomes more sensitive as the
number of observed GRBs increases, does not require any knowledge of
the GWB waveforms, is insensitive to the presence of non-Gaussian
detector noise, and does not require statistical independence of the
detectors or knowledge of their correlated noise.  It is thus a
powerful addition to the growing arsenal of analysis techniques aimed
at making gravitational wave detection an astronomical tool.

 
\acknowledgments We gratefully acknowledge the hospitality of the LIGO
Laboratory at Caltech, where the work described here was begun.  LSF
is glad to acknowledge discussions with B.~Barish, who drew attention
to the subtleties of upper limit analyses.  SDM acknowledges a
fruitful discussion with E.~S.~Phinney regarding GRBs.  This work was
supported by National Science Foundation grants PHY93-08728,
PHY95-03084 and PHY98-00111.



\end{document}